\documentclass{revtex4}
\usepackage[spanish]{babel}
\usepackage{graphicx}
\usepackage[latin1]{inputenc}
\usepackage{amsmath}
\usepackage{amssymb}
\usepackage{verbatim}

\begin{document}

\markboth{J. Pardo Vega and H. P\'erez Rojas}{Photon Propagation in
the Casimir Vacuum}

\title{PHOTON PROPAGATION IN THE CASIMIR VACUUM}

\author{\footnotesize JAVIER PARDO VEGA}

\affiliation{Facultad de F\'isica, Universidad de la Habana, Cuba
\\jpardo\_v@ictp.it}

\author{HUGO P\'EREZ ROJAS}

\affiliation{Grupo de F\'isica Te\'orica, Instituto de Cibern\'etica, Matem\'atica y
F\'isica, Cuba\\
hugo@icmf.inf.cu}

\begin{abstract}
A transformation that relates the Minkowskian space of the Quantum
Electrodynamics (QED) vacuum between parallel conducting plates and
the QED vacuum at finite temperature is obtained. From this formal analogy,
the eigenvalues and eigenvectors of the photon self-energy for the
QED vacuum between parallel conducting plates (Casimir vacuum) are
found in an approximation independent form. It leads to two
different physical eigenvalues and three eigenmodes. We also apply
the transformation to derive the low energy photons phase velocity
in the Casimir vacuum from its expression in the QED vacuum at
finite temperature.
\end{abstract}

\maketitle

\section{Introduction}


The propagation of light in  vacuum modified by external conditions
(background fields, finite temperature, boundary conditions)
requires careful study since it may lead to conceptual changes such
as the variation of the speed of light with regard to its value in the usual
homogeneous and isotropic QED
vacuum  \cite{Scharnhorst90,Barton90,Latorre95,Scharnhorst98,Dittrich}.
These modified vacua exhibit new properties such as a change in the
dispersion equations and the energy density, with regard to
the usual QED vacuum, leading to observable phenomena like the
Casimir effect. The Casimir Effect is interesting in many fields of
modern physics. For example, in astrophysics, gravitation and
cosmology, it arises in space-time with a nontrivial topology. The
polarization of  vacuum due to the Casimir effect might be important
for the resolution of the problem of the cosmological
constant  \cite{Bordag-Advances}.

The propagation of soft photons through nontrivial QED vacua has
been deeply studied, mostly within the framework of effective
action. In particular, the investigations about light propagation in
the vacuum between parallel conducting plates have lead to the
so-called Scharnhorst Effect (superluminal phase and group
velocities)  \cite{Scharnhorst90,Barton90}. Latorre, Pascual and
Tarrach proposed a unified formula for the low energy change in the
averaged speed of light valid for the modified vacua, which was
generalized by Dittrich and Gies  \cite{Latorre95,Dittrich}.

The analogy between the free energy in a finite temperature field
theory and the Casimir energy for parallel planes under periodic
conditions was firstly discussed by Toms  \cite{Toms}. On the other
hand, the analogy between the Casimir vacuum and the QED vacuum at
finite temperature under certain approximations was firstly pointed
out by Barton  \cite{Barton90} and later by Latorre et al.  \cite{Latorre95}, and by E.
Rodriguez Querts and one of the present
authors (H.P.R.)  \cite{Hugo-Elizabeth}. In the last paper it is
pointed out among other facts, a correspondence between Casimir
vacuum and blackbody radiation energy densities, stemming from the
fact that both problems involve the breaking of a space-time
symmetry  by a constant vector.

In the present paper, we exploit in a more advanced way the
transformation and correspondence between  Casimir vacuum and  QED
vacuum at finite temperature. We obtain the eigenvalues and
eigenvectors of the photon self-energy in an approximation
independent form. Based on this analogy we obtain the known
expression for the low energy photons phase velocity in the Casimir
vacuum at two-loop level and   suggest an alternative interpretation
to the resulting dispersion law.

\section{Correspondence between Thermal and Casimir Vacua}

In the imaginary time formalism of temperature, it is used Euclidean
Field Theory, considered as an analytic continuation from
Minkowskian space by the transformation in the time component of
four-vectors $ct\rightarrow -i\tau$ ($x_0\rightarrow -ix_4$). Then
physical quantities are such that they are not translational
invariant along $\tau$, (but only under the finite shift $\tau \pm
\beta$). In other words Euclidean "time" $\tau$ is restricted to the
interval $[0,\beta]$, where $\beta=\frac{1}{T}$ where $T$ is the
temperature. According to it, fourth components of boson and fermion
fields must satisfy periodic and antiperiodic boundary conditions
with period $\beta$ respectively.

In what follows we will call \emph{thermal vacuum} to a medium at
temperature $T$ where a particle, for instance a photon, propagates
in a background of electrons, positrons and other photons at
temperature $T$. At the tree level, it corresponds to blackbody
radiation.

 For the boson propagator we have
\begin{equation}
D_{\mu\nu}^T(\tau-\beta,\vec{x};\tau',\vec{x}')=D_{\mu\nu}^T(\tau,\vec{x};\tau',\vec{x}'),\:\text{with
} \tau, \tau' \in[0,\beta]
\end{equation}
\noindent and for the fermion propagator
\begin{equation}
S_{\xi\eta}^T(\tau-\beta,\vec{x};\tau',\vec{x}')=-S_{\xi\eta}^T(\tau,\vec{x};\tau',\vec{x}'),\:\text{with
} \tau, \tau'\in[0,\beta].
\end{equation}
\noindent Here $\xi$ and $\eta$ are spinor indices and similar
equations are valid for shifting $\tau'$ in $\beta$. The imposition
of periodical conditions over one coordinate implies that its
conjugate variable in Fourier space takes discrete values.
Therefore, Matsubara discrete frequencies appear in momentum space.

Now, consider the vacuum between two infinitely extended, ideally
conducting and neutral plates parallel to the $1,2$ plane at $x_3=0$
and $x_3=a$, where $a$ is the distance between plates. The
electromagnetic field tensor obeys the following boundary conditions
\begin{equation}
n^\mu \tilde{F}_{\mu\nu}(x)|_{x_3=0,a}=0,\label{fun=0}
\end{equation}
\noindent where $n^\mu=(0,0,0,1)$ is the four-vector whose spatial
part is the normal vector of the plates and
$\tilde{F}_{\mu\nu}=\frac{1}{2}\varepsilon_{\mu\nu\beta\gamma}F^{\beta\gamma}$
is the dual field strength tensor. We will use the metric
$g=\text{diag}(-,+,+,+)$ and we will take $\hbar=c=1$ throughout the
paper.

The photon propagator satisfying the boundary conditions
(\ref{fun=0}) has been computed by Bordag et. al. and has been
widely used by several authors  \cite{Bordag85}. Its expression is
composed by the sum of two terms: the free propagator and a
gauge-independent modification due to the boundaries. Let us make
some reasonable approximations. The experimentally realized
distances between plates are very large in comparison with the
Compton wavelength of the electron. So, the limit $ma\gg 1$ holds
for all realistic situations, where $m$ is the electron mass. We
will also consider that the external electromagnetic field vanishes
near the plates. Therefore, in the momentum
space the limit $ka\gg1$ can be
taken  \cite{Scharnhorst90,Barton90,Latorre95,Dittrich}. Under the approximations
$ma\gg1$ and $ka\gg1$ the photon propagator given by Bordag et. al. leads to the
operative propagator derived by Latorre et. al., which corresponds to the photon
propagator under periodic boundary conditions with period $2a$ for the
spatial coordinate $x_3$  \cite{Latorre95}.

The imposition of boundary conditions on the Dirac field would alter
the fermion propagator, but these changes decrease exponentially
with distance from the plates (on a scale set by the Compton
wavelength) an here we shall neglect
them  \cite{Scharnhorst90,Barton90,Latorre95,Bordag85}. Likewise, for
cold heat baths ($T\ll m$), we can neglect the thermal contribution
due to massive virtual particles which is always exponentially
suppressed  \cite{Dittrich}.

Then, the QED vacuum at low temperature and the vacuum between
parallel conducting plates in the limit considered ($ma\gg 1$,
$ka\gg 1$) are formally analogous, the last one reduces to impose
boundary conditions on photons with period $2a$. Note that, in the
Euclidean space, to do the transformation between the thermal and
Casimir vacua it is enough to exchange the third and fourth
components. Thus, the following transformation relates the
Minkowskian spaces of both vacua:
\begin{align}
x_0&\rightarrow -ix_3 \label{t0}\\
x_1&\rightarrow x_1   \label{t1}\\
x_2&\rightarrow x_2   \label{t2}\\
x_3&\rightarrow ix_0  \label{t3}.
\end{align}
\noindent Here, it is done an analytic continuation in the
components zero and three and the parameters $\beta$ and $2a$ are
identified. We will denote the transformation (\ref{t0}-\ref{t3}) as
$\mathcal{T}$.

\section{Tensor Structure of the photon self-energy in the Casimir Vacuum}

The exact photon propagator $D_{\mu\nu}$ can be expressed through
the free photon propagator and the photon photon self-energy
$\Pi_{\mu\nu}$. To find the Bose-excitation spectrum one must solve
$D_{\mu\nu}^{-1}a^{\nu}=0$
for the excitation field $a^\nu$. Simultaneously the poles of the
Green's function $D_{\mu\nu}$ are found. To fulfill this, it is
convenient to diagonalize the photon self-energy  \cite{Hugo79}.
Then, the Green's function is represented as
\begin{equation}
D_{\mu\nu}(k)=\sum_{i=1}^4 \frac{1}{k^2-\kappa_i(k)}\frac{
b_{\mu}^{(i)} b_\nu^{*(i)} } {b_{\xi}^{(i)} b_{ \:}^{*\xi (i)} },
\end{equation}
\noindent where $\kappa_i$ are eigenvalues and $b_\mu^{(i)}$ are
eigenvectors of the photon self-energy, and the vector potential of
the excitation is represented as
\begin{equation}
a_\mu (k)=\sum_{i=1}^4 \delta(k^2-\kappa_i(k))b_\mu^{(i)}(k).
\end{equation}

Accordingly, to find the dispersion law for the wave whose vector
potential is $b_\mu^{(i)}(k)$ one must solve the equation
$
k^2=\kappa_i(k).
$
One of the eigenvalues, say, with $i=4$, is zero $\kappa_4=0$,
$b_\mu^{(4)}=k_\mu$. This wave is purely longitudinal in the
four-dimensional sense and the electromagnetic field strength for it
is zero. So, it is enough to solve three dispersion equations.

Using the analogy with the thermal vacuum, the polarization tensor
for the Casimir vacuum can be expressed in the Euclidean space as
\begin{align}
\Pi_{\mu\nu}&=\left(\delta_{\mu\nu}-\frac{k_\mu
k_\nu}{\tilde{k}^2}\right)A(\tilde{k}^2,k_3^2)+ \Pi_{33}\frac{k_\mu
k_\nu}{\tilde{k}^2}\frac{k_3^2}{\tilde{k}^2}\label{Pimunu}\\
\Pi_{\mu3}&=\Pi_{3\mu}=-\Pi_{33}\frac{k_\mu k_3}{\tilde{k}^2}\;
,\:\text{where } \mu,\nu=1,2,4.\label{Pi3nu}
\end{align}
\noindent The Eqs. (\ref{Pimunu}) and (\ref{Pi3nu}) have been
obtained based on the structure of the photon self-energy for the
thermal vacuum in the Euclidean space proposed by
Fradkin  \cite{Fradkin}. This structure can be obtained by means of
symmetry considerations.

The eigenvalues and eigenvectors of the photon self-energy for the
Casimir vacuum in the Minkowskian space are
\begin{align}
\kappa_1&=A & b^{(1)}&=(k_1,k_0,0,0)\\
\kappa_2&=A & b^{(2)}&=(0,k_2,k_0,0)\\
\kappa_3&=\Pi_{33}\frac{\tilde{k}^2+k_3^2}{\tilde{k}^2} &
b^{(3)}&=(k_0k_3,k_1k_3,k_2k_3,-\tilde{k}^2).
\end{align}
\noindent The first and second modes are analogous to the transverse
modes of the thermal vacuum, and the third mode is analogous to the
longitudinal mode. For all modes, gauge invariance is preserved. The
polarization directions of electric $\vec{e}\,^{(1,2,3)}$ and
magnetic $\vec{h}^{(1,2,3)}$ fields in the waves with
four-potentials $b_\mu^{(1,2,3)}$ are
\begin{align}
\vec{e}\,^{(1)}&=-i(k_1^2-k_0^2,k_1k_2,k_1k_3)&
\vec{h}^{(1)}&=ik_0(0,k_3,-k_2)\label{direcc1}\\
\vec{e}\,^{(2)}&=-i(k_1k_2,k_2^2-k_0^2,k_2k_3)&
\vec{h}^{(2)}&=ik_0 (-k_3,0,k_1)\label{direcc2}\\
\vec{e}\,^{(3)}&=-ik_0 k^2(0,0,1) &
\vec{h}^{(3)}&=ik^2(-k_2,k_1,0)\label{direcc3}.
\end{align}

For the first and second modes, the electric field is orthogonal to
the vector $(k_1k_3,k_2k_3,-\tilde{k}^2)$ and the magnetic field is
orthogonal to $\vec{k}$. Moreover, in both cases the electric field
has a small component along the direction of $\vec{k}$:
$\vec{e}\,^{(1)}\cdot\vec{k}=-ik_1k^2$ and
$\vec{e}\,^{(2)}\cdot\vec{k}=-ik_2k^2$, which vanishes on the light
cone mass shell $k^2=0$ and for propagation perpendicular to the
plates. For the third mode, the magnetic field is along the
3-direction and the electric field is orthogonal to $\vec{k}$,
however, both fields vanish on the light cone. The particular cases
of propagation parallel and perpendicular to the plates have special
interest and can be easily discussed from Eqs.
(\ref{direcc1}-\ref{direcc3}). E.g., for propagation perpendicular
to the plates, for the third mode $\vec{e}\,^{(3)}||\vec{k}$ and the
magnetic field is zero. Whereas, for propagation parallel to the
plates, from the point of view of symmetry, the electric field
corresponding to the first and second modes can have a small
component along the direction of propagation which vanishes on the
light cone.

Finally, the dispersion equation for the first and second modes is
the same: $k^2-A(\tilde k^2,k_3^2)=0$ and for the third mode is
$\tilde k^2-\Pi_{33}(\tilde k^2,k_3^2)=0$. The results of this
section are general, determined by the symmetry of the system and
independent of the order of approximation in the perturbative loop
expansion. Let us use the transformation $\mathcal{T}$ to obtain the
known dispersion relation for the first and second modes of the
Casimir vacuum from the dispersion relation for transverse modes of
the thermal vacuum. The dispersion relation for the third mode of
Casimir vacuum has not been computed in the literature.

\section{Dispersion Relations}

In the limit $T\ll m$,
thermal one loop effects are exponentially suppressed by the
electron mass. Similarly, in the Casimir vacuum, one loop
contributions to the dispersion relations due to the boundaries are
exponentially suppressed. However, the two-loop contribution
involves a virtual photon within the fermion loop. In the low
temperature regime the two loop contribution exceeds the influence
of the one loop part due to the thermal excitation of the internal
photon. Correspondingly, in order to obtain the radiative
corrections to the dispersion relations in the Casimir vacuum one
has to consider the two loop contribution.

\noindent By applying the transformation $\mathcal{T}$  to the low energy photons
dispersion equation for transverse modes of the thermal vacuum \cite{Barton90}, we
obtain the dispersion equation for modes one and two in the Casimir vacuum
\begin{equation}
\omega^2= k^2\left(1 + \alpha^2  \frac{11 \pi^2 }{4050}\frac{1}{a^4
m^4} cos^2 \theta \right),
\label{efec}
\end{equation}
\noindent
\noindent where $\theta$ is the angle
between the direction of propagation and the normal to the plates. Hence, computing
the phase velocity we arrive to the Scharnhorst's result
$v=\frac{\omega}{|\vec{k}|}=1+\frac{11\pi^2}{8100}\alpha^2
\frac{1}{a^4m^4}\cos^2\theta$.
The phase
and group velocities coincide and, according to the standard
interpretation, are greater that $c$ for propagation perpendicular
to the plates \cite{Scharnhorst98}.

However,  the breaking of translational invariance in Casimir
vacuum,  leads to non-conservation of momentum along th 3-direction.
Thus we suggest an alternative interpretation of the dispersion
relation of the first and second modes by understanding (\ref{efec})
as meaning
 an effective photon momentum in the 3-direction $k_3'=\left(
1+\frac{11\pi^2}{8100}\alpha^2 \frac{1}{a^4m^4}\right)k_3$,
conserving the dispersion relation
$\omega=|\vec{k}|$. 
This last interpretation is consistent with the fact that the second
order corrections to the Casimir force are repulsive.

\section{Conclusions}

We found an explicit transformation ($\mathcal{T}$) relating
Minkowskian thermal vacuum and Casimir vacuum spaces. Based on this
transformation, the structural properties of the photon self-energy
and the polarizations of eigenmodes were described. For eigenmodes
one and two the electric field has small component along $\vec{k}$
which vanishes on the light cone and the magnetic field is
orthogonal to $\vec{k}$; while both fields in the third mode vanish
on the light cone. The dispersion relation for low energy photons
for the first and second modes in Casimir vacuum within the two loop
approximation can be re-obtained based on the analogy with the
thermal vacuum and we proposed an alternative interpretation keeping
photons on the light cone.

\section*{Acknowledgements}
The  authors thank the OEA-ICTP for its support through NET-35.


\begin{thebibliography}{0}

\bibitem{Scharnhorst90} K. Scharnhorst, \emph{Phys. Lett. B} \textbf{236} (1990) 354.

\bibitem{Barton90} G. Barton,  \emph{Phys. Lett. B} \textbf{237} (1990) 559.

\bibitem{Latorre95} J. Latorre, P. Pascual and R. Tarrach, \emph{Nucl. Phys. B} \textbf{437} (1995) 60.

\bibitem{Scharnhorst98} K. Scharnhorst, ``The velocities of light in modified QED vacua'',
arXiv:hep-th/9810221v2, 1998.

\bibitem{Dittrich} W. Dittrich and H. Gies, {\it Phys. Rev. D} \textbf{58} (1998) 025004.

\bibitem{Bordag-Advances} M. Bordag, G.L. Klimchitskaya, U.
Mohideen, V.M. Mostepanenko, {\it Advances in the Casimir Effect}
(Oxford University Press, USA, 2009).

\bibitem{Toms} D.J. Toms, {\it  Phys. Rev. D} {\bf 21} (1980) 928.

\bibitem{Hugo-Elizabeth} H. P\'erez Rojas and E. Rodr\'iguez Querts,
{\it Int. J. Mod. Phys. A} {\bf 21} (2006) 3761.

\bibitem{Bordag85} M. Bordag, D. Robaschik and E. Wieczorek, \emph{Ann. Phys.} \textbf{165} (1985)
192.

\bibitem{Fradkin} E. S. Fradkin, in {\it Quantum Field Theory and
Hydrodynamics --- Proc. P.N. Lebedev Phys. Inst.}, ed. D. V.
Strobel'tsyn (Consultants Bureau, New York, 1967).

\bibitem{Hugo79} H. P\'erez Rojas and A. E. Shabad, \emph{Ann. Phys.}
\textbf{121} (1979) 432.



\end{thebibliography}
\end{document}